\documentclass[sigconf]{acmart}

\setcopyright{none}
\AtBeginDocument{%
  }

\usepackage[utf8]{inputenc}
\usepackage{soul}
\usepackage{xspace}
\usepackage{textcomp}
\usepackage{listings}
\usepackage{multirow}
\usepackage{colortbl}
\usepackage{lscape} 
\usepackage{rotating}
\usepackage{url} 
\usepackage{pifont}
\usepackage{framed}

\definecolor{jbblue}{HTML}{0573EB}
\definecolor{jbgreen}{HTML}{1D8574}
\definecolor{jborange}{HTML}{C15703}
\definecolor{jbpurple}{HTML}{6b57ff}
\definecolor{jbpink}{HTML}{E90067}
\definecolor{lightgray}{gray}{0.9} 
\definecolor{codegreen}{rgb}{0,0.6,0}
\definecolor{codegray}{rgb}{0.5,0.5,0.5}
\definecolor{codepurple}{rgb}{0.58,0,0.82}
\definecolor{backcolour}{rgb}{0.95,0.95,0.92}
\definecolor{APA_stats}{RGB}{100, 100, 120}
\definecolor{main}{HTML}{5989cf}   
\definecolor{sub}{HTML}{cde4ff}     
\colorlet{shadecolor}{sub}

\newenvironment{RQanswer}{%
  \MakeFramed{\advance\hsize-\width \noindent \FrameRestore}}%
{\endMakeFramed}

\newcommand{\Topic}[2]{\textcolor{#1}{\textbf{#2}}}

\newcommand{\company}{JetBrains\xspace}
\usepackage[english]{babel}
\usepackage[autostyle, english = american]{csquotes}
\MakeOuterQuote{"}

\copyrightyear{2026}
\acmYear{2026}
\setcopyright{cc}
\setcctype{by}
\acmConference[ICSE-SEIP '26]{2026 IEEE/ACM 48th International Conference on Software Engineering}{April 12--18, 2026}{Rio de Janeiro, Brazil}
\acmBooktitle{2026 IEEE/ACM 48th International Conference on Software Engineering (ICSE-SEIP '26), April 12--18, 2026, Rio de Janeiro, Brazil}
\acmPrice{}
\acmDOI{10.1145/3786583.3786883}
\acmISBN{979-8-4007-2426-8/2026/04}

\begin{document}

\title{Developer Needs and Feasible Features for AI Assistants in IDEs}


\author{Agnia Sergeyuk}
\authornote{Both authors contributed equally to this research.}
\affiliation{%
  \institution{JetBrains Research}
  \city{Belgrade}
  \country{Serbia}
}
\affiliation{%
  \institution{Delft University of Technology}
  \city{Delft}
  \country{Netherlands}
}
\email{a.serheyuk@tudelft.nl}

\author{Ekaterina Koshchenko}
\authornotemark[1]

\affiliation{%
  \institution{JetBrains Research}
  \city{Amsterdam}
  \country{Netherlands}
}
\email{ekaterina.koshchenko@jetbrains.com}

\author{Ilya Zakharov}

\affiliation{%
  \institution{JetBrains Research}
  \city{Belgrade}
  \country{Serbia}
}
\email{ilia.zaharov@jetbrains.com}

\author{Timofey Bryksin}

\affiliation{%
  \institution{JetBrains Research}
  \city{Limassol}
  \country{Cyprus}
}
\email{timofey.bryksin@jetbrains.com}

\author{Maliheh Izadi}

\affiliation{%
  \institution{Delft University of Technology}
  \city{Delft}
  \country{Netherlands}
}
\email{m.izadi@tudelft.nl}

\renewcommand{\shortauthors}{Sergeyuk et al.}

\begin{abstract}
  Despite the increasing presence of AI assistants in Integrated Development Environments (IDEs), it remains unclear what different groups of developers actually need from these tools and which features are likely to be implemented in practice. To investigate this gap, we conducted a two-phase study. First, we interviewed 35 professional developers from three user groups (Adopters, Churners, and Non-Users) to uncover unmet needs and expectations. Our analysis revealed five key areas of need distinctly distributed across practitioners' groups: \textit{Technology Improvement}, \textit{Interaction}, and \textit{Customization}, as well as \textit{Simplifying Skill Building}, and \textit{Programming Tasks}. We then examined the feasibility of addressing selected needs through an internal prediction market involving 102 practitioners. The results demonstrate a strong alignment between the developers' needs and the practitioners' judgment for features focused on implementation and context awareness. However, features related to proactivity and maintenance remain both underestimated and technically unaddressed. Our findings reveal gaps in current AI support and provide practical directions for developing more effective and sustainable in-IDE AI systems.
\end{abstract}




\maketitle

\section{INTRODUCTION}

The increasing integration of Artificial Intelligence (AI) tools in software development is transforming the way developers interact with their Integrated Development Environments (IDEs)~\cite{marron2024new, izadi2024language, Semenkin2024Context}. These AI-powered assistants facilitate tasks like code generation, debugging, and optimization with the goal of boosting productivity and reducing the cognitive load associated with repetitive programming tasks~\cite{githubcopilot, jetbrainsai, cursorai, sergeyuk2025human}. As AI-based systems evolve rapidly, teams that build tools for tech creators face two key challenges: understanding how coding specialists actually use AI inside their IDEs and deciding which features are worth investing in. 

However, actionable insights into in-IDE human-AI experience (HAX) are still limited. While recent studies have explored developers' general expectations for AI-assisted tools, few focus specifically on in-IDE interactions, particularly the successes and challenges developers face and their needs with these tools~\cite{sergeyuk2025human}. Even fewer studies consider differences across user groups: those who adopted AI into their workflow, those who abandoned it, and those who have never tried it. These distinctions are important for informing development and marketing strategies~\cite{rogers2014diffusion}. Moreover, these studies often focus on usage patterns or user perceptions without connecting them to concrete feature development decisions. As a result, we know little about how specific user needs translate into buildable and valuable AI features.

To address existing knowledge gaps, we conducted a multi-method study in two phases: qualitative semi-structured interview~\cite{kallio2016systematic} and prediction market~\cite{wolfers2004prediction}. First, we interviewed 35 tech creators across three usage groups (Adopters, Churners, and Non-users) to gain qualitative insights into their experiences, challenges, and unmet needs with current AI features. This phase reveals unmet needs and expectations across five areas: Technology Improvement, Interaction, Customization, Simplifying Skill Building, and Programming Tasks. Yet, understanding user needs alone does not directly translate into feasible product development strategies. Prioritizing feature development requires additional insight into both technical feasibility and organizational readiness. To achieve this, we complemented interviews with a prediction market. Participants predicted which features would be implemented, using collective judgment to estimate outcomes.
We asked 102 internal practitioners to predict the likelihood of implementing 12 user-desired features. This allows us to integrate user-driven insights with practical considerations from the very people who shape our product development.
Our results reveal that internal practitioners correctly anticipate the implementation of features around AI in Implementation and Context-aware Technology. In contrast, features requiring proactive behavior and maintenance support remain undervalued and underdelivered. These findings highlight a misalignment between developer needs and current feasibility assessments, pointing to concrete opportunities for more helpful and forward-looking AI assistance in IDEs.

With this work, we aim to provide industry practitioners and researchers with structured insights to guide product decisions. Specifically, offer:
\begin{itemize}   
    \item A categorization of user needs for in-IDE HAX to outline requirements to make AI tools more useful;
    \item A comparison of perspectives across Adopters, Churners, and Non-Users, revealing group-specific concerns that influence adoption and retention, such as trust and technical issues among Churners and ethical reservations and skill development concerns among Non-Users.
    \item A mapping of the externally voiced needs to internal perceptions of feasibility, identifying areas of alignment (e.g., implementation and context-awareness) and areas of mismatch (e.g., proactivity and maintenance support).
\end{itemize}

We are the first to triangulate detailed user needs with internal feasibility estimates using prediction markets. This approach enables organizations to better prioritize AI features grounded in what users want and what teams can build.
\section{BACKGROUND}

AI-powered programming assistants have become increasingly common, with tools such as GitHub Copilot, Cursor, and JetBrains AI integrated into developer workflows~\cite{githubcopilot,jetbrainsai,cursorai}. While their technical capabilities continue to evolve, research into how developers experience and evaluate such tools has also grown. 

Recent studies have investigated what professional developers need and expect from AI-powered programming assistants~\cite{liang2024large,Mcnutt2023Design,Liu2024Empirical,sergeyuk2024using,vaithilingam2022expectation,zhang2023demystifying,Wang2023How}. Across contexts, developers consistently seek tools that reduce repetitive effort and provide relevant suggestions without introducing new debugging burdens~\cite{liang2024large,zhang2023demystifying,vaithilingam2022expectation}. Effective assistance is expected to align with the current task, understand code context, and support natural language interaction~\cite{liang2024large,Liu2024Empirical,van2023enriching}. Transparency and control are also emphasized, including the ability to steer and correct outputs and control context~\cite{Mcnutt2023Design,liang2024large}. Several studies highlight that experienced users tend to expect more than raw generation~\cite{Liu2024Empirical,Mcnutt2023Design}. For example, tools are valued when they adapt to user intent and surface relevant, task-specific information~\cite{liang2024large,Liu2024Empirical,de2024transformer}. At the same time, integration challenges persist. Developers often report friction when tools disrupt their workflow or fail to fit specific environments~\cite{sergeyuk2024using,vaithilingam2022expectation}. Moreover, non-functional requirements such as performance and scalability may be harder to maintain~\cite{sergeyuk2024using}. These insights point to the need for assistants that can flexibly support diverse roles, offer customization, and operate within existing development processes~\cite{Mcnutt2023Design,sergeyuk2024using,liang2024large}.

However, most studies examine developer interaction with AI in broad settings, without addressing the unique characteristics of IDE-based workflows. This distinction is important. According to large-scale industry reports, up to 97\% of developers rely on IDEs for software tasks~\cite{stackoverflow2023}. In-IDE AI assistants operate within a structured environment that offers richer contextual signals such as file scope, project layout, and cursor history. These tools have the potential to deliver more context-aware support while minimizing context switching. Yet, their design introduces new constraints, including the need to remain non-intrusive and responsive in real time. Therefore, in-IDE AI tools shape distinct user needs, expectations, and challenges that warrant focused investigation.

Moreover, existing studies tend to treat developers as a single user group, overlooking differences across adoption stages~\cite{sergeyuk2025human}. However, industry practice suggests that design and feature prioritization must account for the heterogeneous needs of distinct groups: Adopters, Churners, and Non-Users~\cite{rogers2014diffusion, amershi2019guidelines}. These groups not only engage (or not) with AI tools differently, but also hold divergent mental models, trust levels, and expectations, which may lead to different product requirements or marketing strategies. Recognizing these ``personas'' is necessary to design AI assistants that support retention, re-engagement, or onboarding.

At the same time, even when user needs are clearly articulated, translating them into product decisions remains a challenge. Feature roadmaps must also reflect internal constraints such as technical feasibility, resource allocation, and infrastructure readiness. Traditional prioritization techniques, such as expert panels and surveys, often fall short by providing limited collective insights or biased individual opinions~\cite{avella2016delphi}. An alternative approach, prediction markets, has been shown to effectively aggregate collective expert judgment to forecast the likelihood of specific outcomes~\cite{wolfers2004prediction}. 

A prediction market is a structured method wherein participants place bets on future events, effectively leveraging the ``wisdom of the crowd'' to estimate probabilities accurately~\cite{wolfers2004prediction, arrow2008promise}. Previously, prediction markets have been successfully used to forecast a broad range of outcomes from presidential votes~\cite{Berg2014IEM}, to replicability of scientific research~\cite{Dreber2015ReproducibilityPM}, and the future of real products on the market~\cite{Bryant1994HPInkjet}. Google has been a pioneer in using internal prediction markets for software engineering and product development decisions. Research by ~\citet{cowgill2015corporate} analyzed internal prediction markets of several companies including Google, finding they produced more accurate forecasts than experts about company-specific events like product launch dates and business metrics. While prediction markets have shown promise in corporate technology settings, academic research on their application to software engineering research and practice remains limited. This gap represents a significant opportunity for HCI and software engineering research to examine how prediction markets can be adapted for the specific technical and social contexts of modern software development.

Thus, this paper combines qualitative interviews and prediction market, integrating rich user insights with realistic assessments from internal stakeholders. This dual perspective enables a grounded understanding of both user needs and internal feasibility, offering a practical contribution to the design and deployment of in-IDE AI assistants. We address the following research questions:
\begin{itemize}
    \item \textbf{RQ1.} What are developers' needs when using in-IDE AI tools?

 \item \textbf{RQ2.} How do these needs differ between Adopters, Churners, and Non-Users?

 \item \textbf{RQ3.} To what extent do practitioners acknowledge and correctly anticipate which AI features inspired by user needs are likely to be implemented?
\end{itemize}

\section{UNDERSTANDING DEVELOPER NEEDS} \label{sec:userstudy}

In this section, we examine how developers experience AI inside their IDEs when coding. Our goal is to provide a structured understanding of what developers need from AI assistance, what challenges they face, and how these vary across adoption groups. We address \textbf{RQ1} and \textbf{RQ2} through qualitative interviews.

\subsection{METHOD} \label{sec:method}
The study was conducted in May-June 2024 in line with our institutional ethical standards, adhering to the values and guidelines outlined in the ICC/ESOMAR International Code~\cite{iccesomar}. 

\subsubsection{\textbf{Recruitment}}
We recruited participants from users of \company, a software engineering company that also develops AI-based tools for coding, to ensure a consistent experience with development environments. Note that with that, participants could use any in-IDE AI assistant. Invitations were sent via email to consenting subscribers and promoted through the authors' social media.

Respondents completed a qualification survey. It was designed to assign them to one of three groups based on a question about AI tool experience: current users were classified as \textit{Adopters}, former users as \textit{Churners}, and those who never used such tools as \textit{Non-users}. We chose these three groups to capture various experiences with AI tools in IDEs~\cite{rogers2014diffusion}. These groups are referred to using the prefixes A, C, and N, respectively, in participant identifiers throughout the paper. In the survey, we collected data on job roles, experience levels, IDEs used, AI tools tried for coding, and duration of AI usage. Out of 380 responses, we iteratively invited participants until data saturation was reached~\cite{HENNINK2022114523}, which means that no new themes or insights emerged from the last interviews. As a result, we conducted 35 interviews: 15 Adopters, 12 Churners, and 8 Non-Users. Most of the participants were developers (32) with overlapping roles: 9 team leads, 8 architects, and 5 DevOps engineers. The sample included 28 seniors, 5 mid-level practitioners, and 2 juniors, with experience ranging from under 5 years (9) to over 11 years (16). Participants chose a reward between a USD 100 Amazon gift card or an equivalent-value pack of all \company products.

\subsubsection{\textbf{Data Collection}}

The interviews followed a semi-structured script developed by the first two authors, reviewed by certified UX researchers at JetBrains, and piloted with four internal participants~\cite{kallio2016systematic}. The script included open-ended questions on AI experience, utilized tools, challenges and successes with in-IDE HAX, unmet needs, desired features, and visions for AI in coding.

All interviews were conducted remotely by one of the first two authors, lasting 71 minutes on average (range: 46-92). Each interview followed a common structure covering challenges, successes, needs, and future expectations, with question framing adapted to each group's AI experience while maintaining cross-group comparability. For Adopters, the questions focused on their ongoing use of AI tools, exploring typical workflows, challenges, and successes. Churners, those who have previously used AI tools but then stopped using them, reflected on their initial adoption and further disengagement, identifying reasons for it and missed opportunities. Non-Users were asked about their reasons for not adopting AI, any potential barriers and possible motivations that could encourage them to begin using these tools. This enabled group-specific depth while maintaining cross-group comparability during analysis.

\subsubsection{\textbf{Data Analysis}}

All interviews were transcribed using Condens\footnote{Condens: a user research repository for analyzing qualitative data. \url{https://condens.io/}}. We applied a qualitative content analysis~\cite{forman2007qualitative,elo2008qualitative} to identify user-reported successes, challenges, unmet needs, and future expectations for in-IDE AI. The first two authors reviewed the first transcript together to agree on the analysis and the code-deriving process. They then went into an iterative inductive process, where they independently assigned \colorbox{gray!20}{\textit{Codes}} to quotes from the transcripts. Eventually, they compiled a list of 161 codes, met again to add or remove codes, and resolved any disagreements until full consensus was reached. The final versions of the codes were then assigned to the corresponding quotes. The codes were then grouped into \textcolor{jbgreen}{Thematic groups} representing key areas of need. These were organized into broader \colorbox{jbgreen!50!}{Topics} that describe the categories of in-IDE HAX needs. For example, the following quote:
\begin{quote}     \addtolength{\leftskip}{-.6cm}     \addtolength{\rightskip}{-.6cm}
\textit{``The most useful thing you could do with AI would be a summary of a Pull Request.'' — N13}
\end{quote}
was coded as \colorbox{gray!20}{``Code Review Support''}, assigned to the group \textcolor{jbgreen}{``Code Optimization''}, under the topic \colorbox{jbgreen!50!}{``Simplifying Programming Tasks''}.
Following established practices, we identified emergent patterns across user groups, recognizing that in qualitative research, the presence and salience of themes matter more than frequency counts.

All materials used for data gathering, together with a full demography table of participants and a summarized codebook, are included in Supplementary Materials~\cite{supp_mat}. Full interview transcripts cannot be shared due to the company's confidentiality policy.

\subsection{FINDINGS}
\begin{figure*}[t]
  \centering
  \includegraphics[width=.85\textwidth]{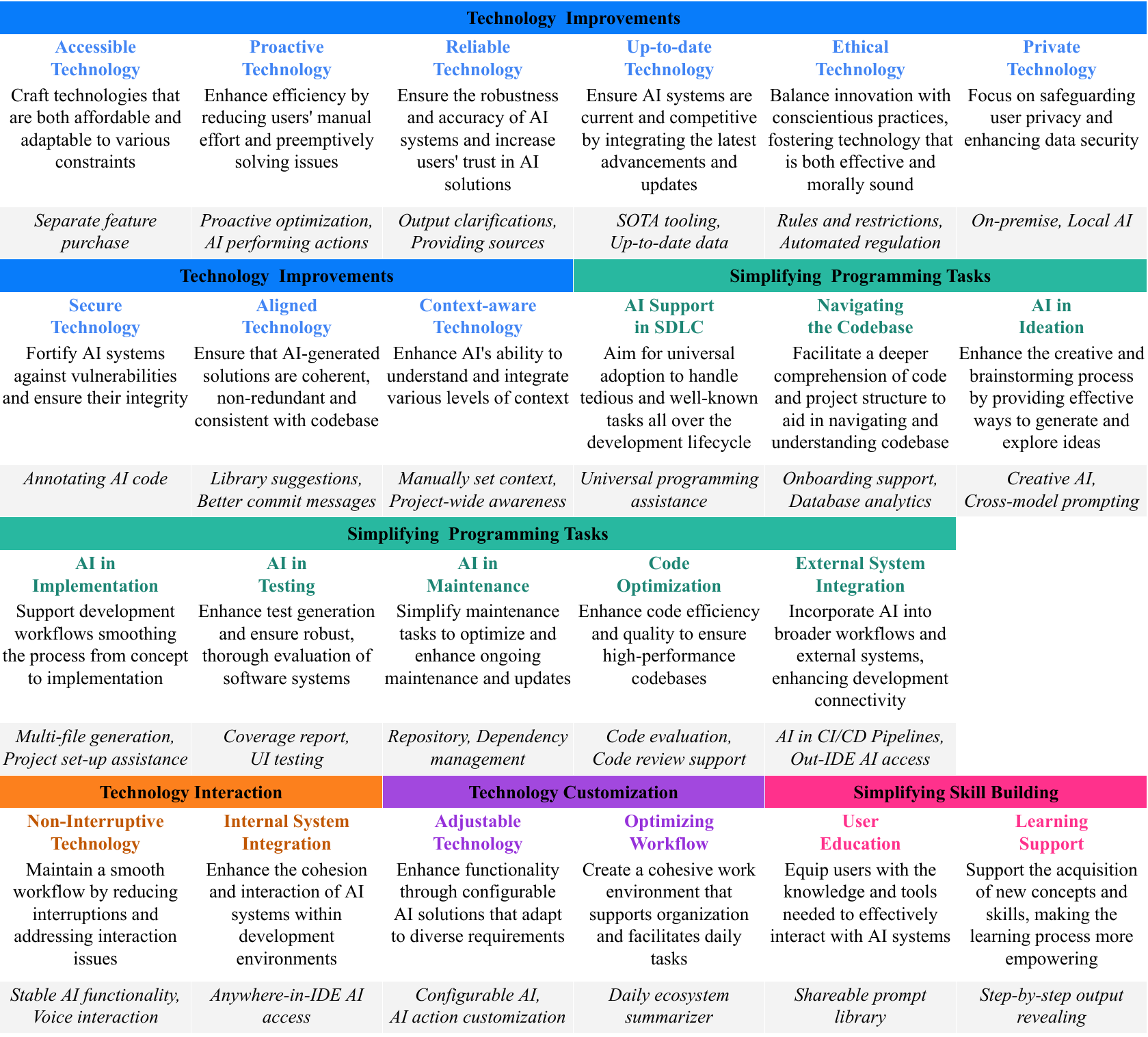}
  \caption{The Categorization of Users' Needs in HAX Within IDEs With Descriptions}.
  \Description{A hierarchical taxonomy diagram presenting 25 user needs for AI assistants in IDEs, organized into five color-coded categories. Each need includes a title, description, and example features. Technology Improvements (orange header) contains nine needs: Accessible Technology (craft affordable and adaptable technologies; examples: separate feature purchase), Proactive Technology (reduce manual effort and preemptively solve issues; examples: proactive optimization, AI performing actions), Reliable Technology (ensure robustness and accuracy to increase trust; examples: output clarifications, providing sources), Up-to-date Technology (integrate latest advancements; examples: SOTA tooling, up-to-date data), Ethical Technology (balance innovation with conscientious practices; examples: rules and restrictions, automated regulation), Private Technology (safeguard privacy and data security; examples: on-premise, local AI), Secure Technology (fortify against vulnerabilities; examples: annotating AI code), Aligned Technology (ensure coherent, non-redundant solutions consistent with codebase; examples: library suggestions, better commit messages), and Context-aware Technology (understand and integrate various context levels; examples: manually set context, project-wide awareness). Simplifying Programming Tasks (purple header) contains five needs: AI Support in SDLC (universal adoption for tedious tasks across development lifecycle; examples: universal programming assistance), Navigating the Codebase (facilitate deeper code comprehension; examples: onboarding support, database analytics), AI in Ideation (enhance creative brainstorming; examples: creative AI, cross-model prompting), AI in Implementation (support development workflows from concept to implementation; examples: multi-file generation, project setup assistance), AI in Testing (enhance test generation and evaluation; examples: coverage report, UI testing), AI in Maintenance (simplify optimization and ongoing updates; examples: repository and dependency management), Code Optimization (enhance code efficiency and quality; examples: code evaluation, code review support), and External System Integration (incorporate AI into broader workflows and CI/CD pipelines; examples: AI in CI/CD pipelines, out-of-IDE AI access). Technology Interaction (pink header) contains two needs: Non-Interruptive Technology (maintain smooth workflow by reducing interruptions; examples: stable AI functionality, voice interaction) and Internal System Integration (enhance cohesion of AI within development environments; examples: anywhere-in-IDE AI access). Technology Customization (teal header) contains two needs: Adjustable Technology (configurable AI adapting to diverse requirements; examples: configurable AI, AI action customization) and Optimizing Workflow (create cohesive work environment supporting daily tasks; examples: daily ecosystem summarizer). Simplifying Skill Building (green header) contains two needs: User Education (equip users with knowledge to effectively interact with AI; examples: shareable prompt library) and Learning Support (support acquisition of new concepts and skills; examples: step-by-step output revealing).}
  \label{fig:full_space}
  \Description{}
\end{figure*}

\subsubsection{\textbf{Categorization of Developer Needs}}

\autoref{fig:full_space}, presents our categorization of developer needs regarding Human-AI Experience within IDEs, based on a thematic analysis of 35 interviews. It includes five overarching topics: Technology Improvements, Simplifying Programming Tasks, Technology Interaction, Technology Customization, and Simplifying Skill Building.

\begin{figure*}[t]
  \centering
  \includegraphics[width=.85\textwidth]{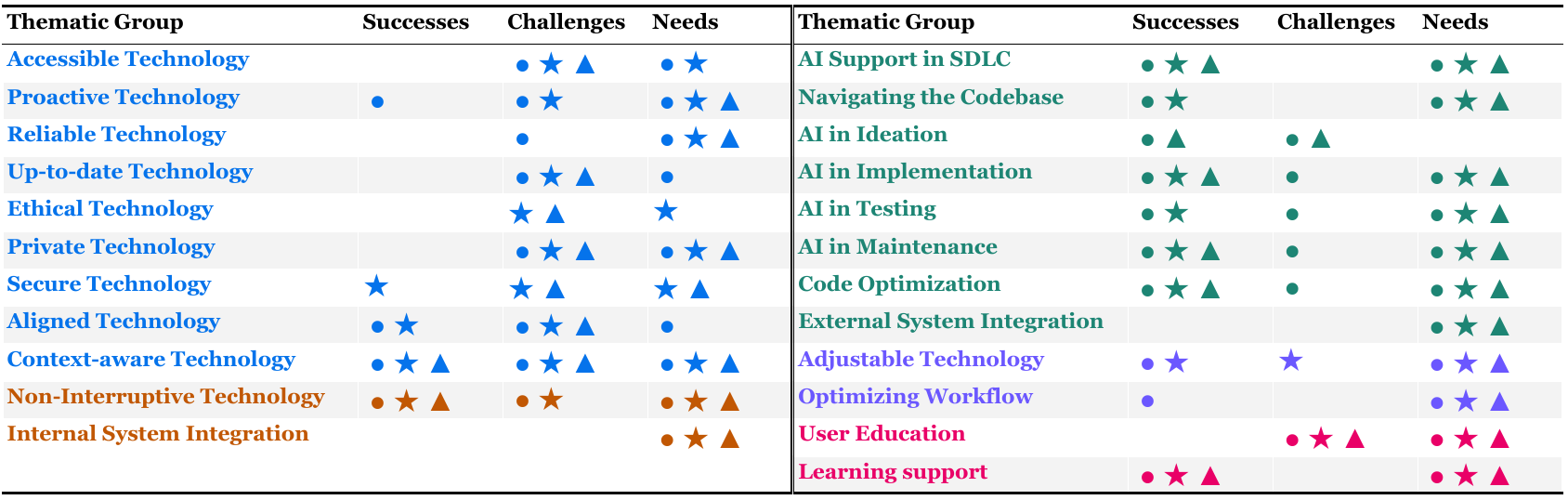}
  \Description{A two-column table showing 25 thematic groups organized by topic area, with columns indicating which participant groups (Adopters, Churners, Non-Users) mentioned each theme as Successes, Challenges, or Needs. The left column contains technology-related themes including Accessible Technology, Proactive Technology, Reliable Technology, Up-to-date Technology, Ethical Technology, Private Technology, Secure Technology, Aligned Technology, Context-aware Technology, Non-Interruptive Technology, and Internal System Integration. The right column contains task and interaction themes including AI Support in SDLC, Navigating the Codebase, AI in Ideation, AI in Implementation, AI in Testing, AI in Maintenance, Code Optimization, External System Integration, Adjustable Technology, Optimizing Workflow, User Education, and Learning Support. Symbols indicate group mentions: filled circles for Adopters, stars for Churners, and triangles for Non-Users. Themes are color-coded by category: orange for Technology Improvement, teal for Interaction, blue for Customization, purple for Programming Tasks, and green for Simplifying Skill Building. The table reveals that Adopters reported more Successes across themes, while Churners and Non-Users more frequently mentioned Challenges and Needs, particularly in areas like Proactive Technology, Context-aware Technology, and AI in Maintenance.}
  
  {\footnotesize \textit{Note:} Symbols represent the participant groups: Adopters (\ding{108}), Churners (\ding{72}), and Non-Users (\ding{115}). The presence of a symbol indicates that group members mentioned the corresponding theme in the associated interview section.}
  
  \caption{Comparative Analysis of Thematic Group Mentions Across User Types.}
  \label{fig:themes_by_users}
\end{figure*}

\paragraph{\colorbox{jbblue!50!}{\textbf{Technology Improvements}}} 

Participants identified several requirements for improvements of AI tooling for coding, beginning with \textcolor{jbblue}{accessibility} across infrastructural, legal, and experiential boundaries. They proposed specific solutions, including natural language programming interfaces, universal language support, and modular pricing models, to reduce barriers in regions with limited connectivity or restrictive regulations. Beyond basic accessibility, participants envisioned AI systems transitioning from reactive assistants to \textcolor{jbblue}{proactive} collaborators. This shift would enable AI to identify potential issues during development:

\begin{quote}     
\addtolength{\leftskip}{-.6cm}     
\addtolength{\rightskip}{-.6cm}  
    \textit{``It would be cool if it could search for problems along the way. It could say `Did you realize you just created an out-of-bounds condition?' ''} --- \textit{A44}
\end{quote}

Such proactive capabilities, however, require fundamental improvements in system \textcolor{jbblue}{reliability} and integration of the \textcolor{jbblue}{latest advancements} in the field. Participants emphasized that unreliable systems create additional verification overhead:

\begin{quote}     
\addtolength{\leftskip}{-.6cm}     
\addtolength{\rightskip}{-.6cm}
    \textit{``If I felt the AI system was really reliable, it would have saved me time because I wouldn't need to double-check and could just apply the right solution right away.''} --- \textit{C159}
\end{quote}

These technical improvements must be accompanied by robust \textcolor{jbblue}{ethical} frameworks. Participants recognized AI's potential and the corresponding responsibility this creates:

\begin{quote}     
\addtolength{\leftskip}{-.6cm}     
\addtolength{\rightskip}{-.6cm}    
    \textit{``AI, like anything with the potential to destroy jobs, carries a huge responsibility. I really hope that the community and organizations invest heavily in getting it right. Otherwise, recovering might be really difficult.''} --- \textit{A37}
\end{quote}

Ethical considerations extend to \textcolor{jbblue}{privacy} and \textcolor{jbblue}{security} requirements. Participants strongly favored on-premise or local AI solutions to ensure data confidentiality and comply with organizational or national restrictions. They emphasized the need for vulnerability prevention and maintaining code integrity, including mechanisms to clearly identify AI-generated code for inspection and verification. Finally, participants stressed that all improvements must support better \textcolor{jbblue}{alignment} with existing development practices. \textcolor{jbblue}{Context awareness} emerged as essential for generating relevant and accurate code. They specifically requested mechanisms to manually set, view, and edit the context provided to AI assistants. Current systems' lack of transparency in context construction creates frustration and undermines user confidence in AI-generated outputs.

\paragraph{\colorbox{jbgreen!50!}{\textbf{Simplifying Programming Tasks}}}

Participants identified AI's primary value in automating routine and repetitive activities across the \textcolor{jbgreen}{software development lifecycle} (SDLC). This automation would enable developers to redirect effort from low-level tasks toward more complex and creative work spanning ideation, implementation, testing, maintenance, and optimization phases. This vision extends beyond simple task automation to AI functioning as an intelligent collaborator:

\begin{quote}     
\addtolength{\leftskip}{-.6cm}    
\addtolength{\rightskip}{-.6cm}
    \textit{``I would say it should be like a colleague, someone who understands what I wrote and can guide or help me where I lack knowledge.''} --- \textit{A26}
\end{quote}

A critical component of this collaborative relationship involves AI-supported \textcolor{jbgreen}{codebase navigation}. Participants particularly valued AI's potential to help newcomers understand existing systems and adhere to established standards:

\begin{quote}     
\addtolength{\leftskip}{-.6cm}     
\addtolength{\rightskip}{-.6cm}
    \textit{``If you gave the AI the company documentation and code and then let it automatically detect what's needed, this would help newcomers follow the company's standards.''} --- \textit{C94}
\end{quote}

Participants also envisioned AI tools that seamlessly connect IDEs with external environments, including Continuous Integration and Continuous Deployment (CI/CD) pipelines, version control platforms, and support ticketing systems.

\begin{quote}     
\addtolength{\leftskip}{-.6cm}     
\addtolength{\rightskip}{-.6cm}
    \textit{``Right now, so many things come through email, then there's the support ticketing system, and, of course, our Git repositories. There could be ways to integrate all these even better.''} --- \textit{C118}
\end{quote}

\paragraph{\colorbox{jborange!50!}{\textbf{Technology Interaction}}}

Participants identified seamless \textcolor{jborange}{internal system integration} as fundamental to an effective AI tool, emphasizing the need for AI to interact efficiently with existing IDE features. Beyond that, participants sought \textcolor{jborange}{uninterrupted} experiences that enable flexible AI interaction modes throughout their development workflows. However, current implementations frequently fall short of this due to technical limitations that disrupt workflow continuity. These disruptions manifest in several forms: slow response times that break the natural rhythm of development, frequent authentication failures that force repeated logins, and functional limitations including copy/paste failures and conversation dead-ends where AI systems become unresponsive. Such technical barriers undermine the seamless integration that participants consider essential for productive AI-assisted development.

\paragraph{\colorbox{jbpurple!50!}{\textbf{Technology Customization}}}

The interviewees stressed the importance of \textcolor{jbpurple}{adjustable} AI systems that offer a high degree of personalization and flexibility, allowing users to retain control as decision-makers. Key features included configurable AI, the ability to choose between different models, and the option to customize AI actions based on specific tasks or workflows. Participants also expressed a desire for task-specific setups that allow AI to adapt to unique user preferences and working styles. To support \textcolor{jbpurple}{workflow optimization}, participants were also interested in tools that could help manage distractions, track time, and summarize daily activities within their development ecosystem. The ability to toggle time tracking, manage task tickets, and maintain a personal prompt library was highlighted as essential for staying organized and improving productivity.

\begin{quote}     \addtolength{\leftskip}{-.6cm}     \addtolength{\rightskip}{-.6cm}
    \textit{``At my job, I have around 20 tickets pending, and sometimes I forget which one I'm currently working on and what to do next, so I need something that can help me stay organized.''} --- \textit{N14}
\end{quote}

\paragraph{\colorbox{jbpink!50!}{\textbf{Simplifying Skill Building}}}

The participants noted that AI tools often have a steep learning curve. They identified the need for better \textcolor{jbpink}{user education}, particularly around effective prompt formulation. A shareable prompt library was seen as a possible valuable resource, as many reported difficulties in crafting prompts that yield useful results. Additionally, the unfamiliar interface of some AI systems contributed to these challenges. Several Non-Users mentioned that the primary reason they had not adopted AI tools was the time required to become proficient.

\begin{quote}     \addtolength{\leftskip}{-.6cm}     \addtolength{\rightskip}{-.6cm}
    \textit{``It's a useful tool if employed correctly. But there is a learning curve. If I figure out how to work with it efficiently, how to set up the right prompts, etc., it should speed up my work.''} --- \textit{N84}
\end{quote}

The participants also emphasized the role of AI as a \textcolor{jbpink}{learning enabler} and mentor, helping developers acquire new skills and understand new concepts. Features such as step-by-step output reveal were considered helpful in fostering long-term skill development through more interactive and adaptive guidance.

\begin{quote}     \addtolength{\leftskip}{-.6cm}     \addtolength{\rightskip}{-.6cm}
    \textit{``This way, you wouldn't have to keep requesting new hints would be more interactive, allowing you to reveal the steps as you go and adjust based on what you need.''} --- \textit{A58}
\end{quote}

\begin{RQanswer}
\textbf{Answer to RQ1.} Developers require AI tools that are advanced, well-integrated, customizable, dedicated to simplifying programming tasks and skill building.
\end{RQanswer}

\subsubsection{\textbf{Comparison of Needs' Themes Among Groups}}

We present a comparative analysis of the Thematic groups among Adopters, Churners, and Non-Users in \autoref{fig:themes_by_users}, focusing on participants' success stories (or anticipated advantages for Non-Users), challenges (or barriers for Non-Users), and needs in the context of in-IDE HAX. The presence of a symbol indicates that at least one participant within that user group mentioned the corresponding theme during their interview. This binary presence/absence approach was chosen to capture the full spectrum of experiences and concerns across adoption stages, as even singular mentions may represent important but underexplored aspects of the user experience that warrant consideration in feature development.

The \textcolor{jbblue}{Context-aware Technology} thematic group is uniquely mentioned by all three groups of participants as a success, a challenge, and a need, indicating the most ambiguous relationship to this aspect of technology. 
\textcolor{jbblue}{Privacy} and \textcolor{jbpink}{User Education} emerge as universal concerns and points of potential and urgent improvement, with all three groups reporting challenges and needs in this regard. 
All three groups of participants mention \textcolor{jbpink}{Learning support} and \textcolor{jbgreen}{AI Support in various stages of SDLC} as success and need, signaling that the evolution of AI tooling is sufficient but still relevant in these areas.  
Notably, \textcolor{jbgreen}{AI in the Ideation} theme is not mentioned by any of the groups as a need, signaling that people are less likely to seek AI support at this stage of SDLC. On the other hand, participants mention \textcolor{jborange}{Internal} and \textcolor{jbgreen}{External System integration} only as a need, implying that this is a valuable attention point for tool builders. 

It is interesting to note that participants usually mention in challenges and needs \colorbox{jbblue!50!}{Technology Improvements}-related themes, while \colorbox{jbgreen!50!}{Simplifying Programming Tasks}-related themes are usually mentioned in successes and needs. This trend might be interpreted as a direction for tool builders to improve technology, making it more advanced, as opposed to widening applications, which are already successful enough.

\subsubsection{\textbf{Comparison of Needs' Codes Among Groups}}
Further, we explore the similarities and differences of participants' 78 specific needs at the Codes' level, represented in Table~\ref{tab:user_group_needs}.

\begin{table}[tbp]
\footnotesize
\setlength{\abovecaptionskip}{3pt} 
  \caption{User group needs across topics.}
  \label{tab:user_group_needs}
  \centering
  \setlength{\tabcolsep}{2pt}
  \renewcommand{\arraystretch}{.95}
  \rowcolors{3}{gray!10}{white}
  \begin{tabular}{@{}p{3.2cm} p{4cm} c c c@{}}
    \toprule
    \textbf{Topic} & \textbf{Code of Need} & \textbf{A} & \textbf{C} & \textbf{N} \\
    \midrule
    \Topic{jbblue}{Accessible Technology} & Polyglot AI &  & \ding{51} &  \\
     & Separate feature purchase & \ding{51} & \ding{51} &  \\
     & Reasonable pricing & \ding{51} & \ding{51} &  \\
     \Topic{jbblue}{Proactive Technology} & Proactive code generation & \ding{51} & \ding{51} & \ding{51} \\
     & Proactive documentation & \ding{51} & \ding{51} &  \\
     & Proactive optimization & \ding{51} & \ding{51} & \ding{51} \\
     & Proactive debugging & \ding{51} & \ding{51} &  \\
     & AI performing actions & \ding{51} & \ding{51} & \ding{51} \\
     & File management & \ding{51} &  &  \\
     & Library management &  & \ding{51} &  \\
      \Topic{jbblue}{Reliable Technology} & Providing sources & \ding{51} &  &  \\
     & Output clarifications & \ding{51} &  &  \\
     & Higher quality & \ding{51} & \ding{51} & \ding{51} \\
    \Topic{jbblue}{Up-To-Date Technology} & Sota tooling & \ding{51} &  &  \\
    \Topic{jbblue}{Ethical Technology} & Ethical technology &  &  & \ding{51} \\
    \Topic{jbblue}{Private Technology} & Privacy &  &  & \ding{51} \\
     & On-premise AI & \ding{51} & \ding{51} & \ding{51} \\
     & Local AI &  & \ding{51} & \ding{51} \\
    \Topic{jbblue}{Secure Technology} & Annotating ai-code &  & \ding{51} & \ding{51} \\
     & Security &  &  & \ding{51} \\
    \Topic{jbblue}{Aligned Technology} & Library suggestions & \ding{51} &  &  \\
     & Generation by example & \ding{51} &  &  \\
    \Topic{jbblue}{Context-Aware Technology} & Contextual awareness & \ding{51} &  & \ding{51} \\
     & Project-wide context awareness & \ding{51} & \ding{51} & \ding{51} \\
     & Business-context awareness & \ding{51} & \ding{51} & \ding{51} \\
     & Manually set context & \ding{51} & \ding{51} & \ding{51} \\
     & Viewable context & \ding{51} & \ding{51} &  \\
     & High level understanding &  & \ding{51} &  \\
    \Topic{jbgreen}{AI Support In SDLC} & Productivity boost & \ding{51} & \ding{51} & \ding{51} \\
     & Universal programming assistance & \ding{51} & \ding{51} & \ding{51} \\
     \Topic{jbgreen}{Navigating The Codebase} & Onboarding support & \ding{51} & \ding{51} &  \\
     & Database analytics &  & \ding{51} & \ding{51} \\
     & Diagram generation & \ding{51} & \ding{51} &  \\
     & Code explanations & \ding{51} &  &  \\
    \Topic{jbgreen}{AI In Implementation} & Project set-up assistance & \ding{51} & \ding{51} &  \\
     & Full stack generation & \ding{51} & \ding{51} &  \\
     & Ui generation & \ding{51} & \ding{51} &  \\
     & Multi-file generation & \ding{51} & \ding{51} & \ding{51} \\
     & Sketch2code &  & \ding{51} &  \\
      \Topic{jbgreen}{AI In Testing} & Coverage report & \ding{51} & \ding{51} &  \\
     & System testing & \ding{51} & \ding{51} & \ding{51} \\
     & Ui testing & \ding{51} &  & \ding{51} \\
     & Error handling &  & \ding{51} & \ding{51} \\
    \Topic{jbgreen}{AI In Maintenance} & Repository management &  & \ding{51} &  \\
     & Database merging &  &  & \ding{51} \\
     & Dependency management & \ding{51} & \ding{51} &  \\
     & Guideline enforcement & \ding{51} &  &  \\
    \Topic{jbgreen}{Code Optimization} & Optimization & \ding{51} &  & \ding{51} \\
     & Code review support & \ding{51} &  & \ding{51} \\
     & Code evaluation & \ding{51} & \ding{51} &  \\
    \Topic{jbgreen}{External System Integration} & Commit management & \ding{51} &  &  \\
     & AI in ci/cd pipelines & \ding{51} & \ding{51} & \ding{51} \\
     & Out-ide AI access & \ding{51} &  &  \\
     & External system integration & \ding{51} & \ding{51} & \ding{51} \\
    \Topic{jborange}{Non Interruptive Technology} & Voice interaction & \ding{51} & \ding{51} &  \\
     & Stable AI functionality &  & \ding{51} &  \\
     & Neuralink & \ding{51} &  &  \\
     & Suggestion preview & \ding{51} & \ding{51} & \ding{51} \\
    \Topic{jborange}{Internal System Integration} & In-line code explanation &  & \ding{51} &  \\
     & Key mappings &  &  & \ding{51} \\
     & AI in code with me & \ding{51} &  &  \\
     & Anywhere-in-ide AI access & \ding{51} & \ding{51} &  \\
     & Internal system integration & \ding{51} &  & \ding{51} \\
    \Topic{jbpurple}{Adjustable Technology} & Personalization & \ding{51} &  &  \\
     & Configurable AI & \ding{51} & \ding{51} & \ding{51} \\
     & Choosing models & \ding{51} & \ding{51} &  \\
     & AI action customization & \ding{51} &  &  \\
    \Topic{jbpurple}{Optimizing Workflow} & Toggleable time tracking &  &  & \ding{51} \\
     & Daily ecosystem summarizer & \ding{51} &  &  \\
     & Distraction management &  &  & \ding{51} \\
     & Ticket management &  & \ding{51} & \ding{51} \\
     & Automation of repetitive tasks & \ding{51} & \ding{51} & \ding{51} \\
     & Proactive workflow recommendations & \ding{51} & \ding{51} &  \\
     \Topic{jbpink}{User Education} & Prompting support & \ding{51} & \ding{51} &  \\
     & Shareable prompt library & \ding{51} &  &  \\
     & User education &  & \ding{51} & \ding{51} \\
    \Topic{jbpink}{Learning Support} & Step-by-step output revealing & \ding{51} &  &  \\
     & Learning support & \ding{51} & \ding{51} & \ding{51} \\
    \bottomrule
  \end{tabular}

  {\footnotesize  \textit{Note: A = Adopter, C = Churner, N = Non-User}}
\end{table}

All three groups expressed needs for comprehensive development support by AI (universal programming assistance, higher quality output, productivity enhancement, multi-file generation, system testing), enhanced tools' context awareness (project-wide and business-context awareness, manually set context), infrastructure flexibility (on-premise AI, CI/CD pipeline integration, external system integration), proactive AI capabilities (code generation, optimization, autonomous actions), and improved user experience (suggestion preview, configurable AI, task automation, learning support). The convergence on these needs suggests they represent core requirements for AI-powered development environments

\textbf{Adopters} demonstrated the broadest range of needs, expressing requirements across all identified categories. This comprehensive need profile reflects their active engagement with AI tools and suggests that continued innovation in advanced features will strengthen retention among this group. Drawing from their relatively extensive AI experience, uniquely identified needs are: 
\begin{itemize}
    \item Advanced tooling and management features (SOTA tooling, File management, Guideline enforcement, Code explanations),
    \item Enhanced alignment and reliability (Personalization, AI action customization, Library suggestions, Generation by example, Providing sources, Output clarifications),
    \item Deeper integration capabilities (Commit management, out-IDE AI access, AI in Code With Me, Neuralink),
    \item Workflow and learning optimization (Daily ecosystem summarizer, Step-by-Step Output Revealing, Shareable Prompt Library).
\end{itemize}

\textbf{Churners} expressed needs that align substantially with Adopters, sharing 36 common requirements across most topics. However, Churners uniquely emphasized:
\begin{itemize}
    \item System stability (Stable AI functionality),
    \item Specific conceptual technical capabilities (Polyglot AI, High-level understanding),
    \item Infrastructure management (Library management, Repository management),
    \item Specialized interface features (In-line Explanation, Sketch2Code).
\end{itemize}
This pattern suggests that while Churners share Adopters' vision for comprehensive AI for code, their focus on stability and conceptual depth indicates that current implementations may lack the robustness and sophistication necessary to sustain engagement. Therefore, implementation quality rather than feature availability drives disengagement decisions.

\textbf{Non-Users} expressed needs in specific areas that represent potential adoption catalysts. Their unique requirements centered on foundational concerns including:
\begin{itemize}
    \item Ethical technology, privacy protection, and security,
    \item Specialized functionality (Database merging, Key mappings),
    \item Productivity control features (Toggleable time tracking, Distraction management).
\end{itemize}
This pattern suggests that Non-Users might prioritize trust-building mechanisms as prerequisites for adoption, indicating that addressing ethical safeguards may be essential for expanding adoption beyond current users.

\begin{RQanswer}
\textbf{Answer to RQ2.} Adopters seek advanced capabilities, Churners require conceptual improvements, and Non-Users need trust-building.
\end{RQanswer}

\section{ESTIMATING FEATURE FEASIBILITY VIA PREDICTION MARKET}

\begin{table}[tbp]
\footnotesize
\setlength{\abovecaptionskip}{3pt}
  \caption{Prediction market results for proposed AI features.}
  \label{tab:tournament_results}
  \centering
  \setlength{\tabcolsep}{4pt} 
  \renewcommand{\arraystretch}{1.0}
  \rowcolors{3}{gray!10}{white}
  \begin{tabular}{@{}p{2.8cm} p{1.2cm} c c r r r@{}}
    \toprule
    \textbf{Feature} & \textbf{Tool} & \textbf{Resolved} & \textbf{Acc.} & \textbf{Prob.} & \textbf{Vol.} & \textbf{Liq.} \\
    \midrule
    \textcolor{jbgreen}{\textbf{Multi-file Generation}} & Any Tool & \ding{51} & \ding{51} & 98\% & 1946 & 570 \\
                                      & JetBrains IDE & \ding{51} & \ding{51} & 92\% & 4288 & 610 \\
    \textcolor{jbblue}{\textbf{Project-wide Context}} & Any Tool & \ding{51} & \ding{51} & 82\% & 440 & 290 \\
                                      & JetBrains IDE & \ding{51} & \ding{55} & 41\% & 2238 & 790 \\
    \textcolor{jbgreen}{\textbf{Sketch2Code}} & Any Tool & \ding{51} & \ding{51} & 90\% & 1019 & 310 \\
                                      & JetBrains IDE & \ding{55} & \ding{51} & 10\% & 500 & 270 \\
    \textcolor{jbblue}{\textbf{Proactive Docs}} & Any Tool & \ding{55} & \ding{51} & 33\% & 878 & 310 \\
                                      & JetBrains IDE & \ding{55} & \ding{51} & 17\% & 1115 & 370 \\
    \textcolor{jbblue}{\textbf{Business Context}} & Any Tool & \ding{51} & \ding{55} & 22\% & 272 & 150 \\
                                      & JetBrains IDE & \ding{55} & \ding{51} & 10\% & 190 & 130 \\
    \textcolor{jbgreen}{\textbf{Real-Time Refactor}} & Any Tool & \ding{55} & \ding{51} & 19\% & 950 & 370 \\
                                      & JetBrains IDE & \ding{55} & \ding{51} & 8\% & 420 & 250 \\
    \textcolor{jbblue}{\textbf{Proactive Optim.}} & Any Tool & \ding{55} & \ding{51} & 19\% & 650 & 310 \\
                                      & JetBrains IDE & \ding{55} & \ding{51} & 8\% & 325 & 210 \\
    \textcolor{jbblue}{\textbf{Proactive Debug}} & Any Tool & \ding{55} & \ding{51} & 6\% & 750 & 210 \\
                                      & JetBrains IDE & \ding{55} & \ding{51} & 8\% & 340 & 150 \\
    \textcolor{jborange}{\textbf{Brain-computer Interface}} & Any Tool & \ding{55} & \ding{51} & 4\% & 1241 & 490 \\
                                      & JetBrains IDE & \ding{55} & \ding{51} & 1\% & 2150 & 510 \\
    \bottomrule
  \end{tabular}
\par\noindent{\footnotesize \textit{Note:} \textbf{Resolved}: feature launched (\ding{51}) or not (\ding{55}). \textbf{Acc.}: prediction correctness. \textbf{Prob.}: final market probability. \textbf{Vol.} and \textbf{Liq.}: participant engagement metrics.}
\end{table}

Having identified developer needs across user groups (\textbf{RQ1, RQ2}), we now turn to \textbf{RQ3}: assessing which user needs and feature directions tool builders consider feasible and valuable and whether this reflects an accurate intuition about how the field is likely to evolve. To answer this, we conducted a prediction market ``tournament'' on JetBrains' internal platform, where participants bet on whether specific AI features would be released within six months. 

All materials used for data gathering at this phase, as well as the full demography table and a summarized report of the prediction market, are in Supplementary materials~\cite{supp_mat}.

\subsection{METHOD}

\subsubsection{\textbf{Materials}} Based on insights from interviews, we distilled an initial list of 38 potential AI-based IDE features~\cite{supp_mat} grounded in ideas and suggestions expressed by participants. This list was reviewed by seven experienced researchers from industry and academia: three specialists in human–AI experience, one assistant professor in computer science, and three senior software engineering researchers. All reviewers had more than five years of professional experience. Reviewers evaluated and discussed asynchronously each feature on three dimensions: existence in current tools, clarity of description, and suitability for prediction market assessment. Through this collaborative process, we selected 9 features for the prediction market. 

These nine features were selected because they are representative of thematic groups that were mentioned by all three user groups (Adopters, Churners, and Non-Users) as both successes and needs. This cross-group presence indicates broad relevance and high potential for user benefit, which in turn justified their perceived feasibility within the short-term horizon of 6 months. In contrast, features from thematic groups such as \textit{AI Support in SDLC} and \textit{Learning Support}, while also appearing in both successes and needs categories, were expressed in ways that were too generic to be translated into concrete and clearly defined feature candidates. These were therefore excluded from the prediction market to ensure that only features with a precise and operational definition were included.

Based on the final list of features, we formulated 18 binary questions for the prediction market. Each question asked whether a given AI-based feature would be released within the next six months, either in JetBrains' IDEs or any other AI-enabled development tool. These questions were published on JetBrains' internal prediction market platform, where employees place virtual bets to estimate the likelihood of future events. All 18 questions were grouped into a single tournament, where cash prizes (\$1500, \$1000, \$500) were to be awarded to the top three predictors on the leader board to incentivize participation.

\subsubsection{\textbf{Participants}} All participants of the prediction market were employed by \company. A total of 102 participants~\cite{supp_mat} across 84 teams placed bets on at least one question during the tournament. The median tenure was 4 years and 10 months. The most common roles included Software Developer (34 participants), Team Lead (20), Product Manager (6), Analyst (6), Quality Assurance (QA) Engineer (5), and Machine Learning (ML) Engineer (4).

\subsubsection{\textbf{Data collection}} The tournament was launched in December 2024. Each participant received 500 virtual coins to place bets on the outcomes of 18 binary questions. For each question, participants could allocate any number of coins to the YES or NO outcome. Before trading, all participants were informed about tournament rules, trading procedures, and strategies for making informed bets. During trading, the current market probability for each question was visible. Bidding closed on June 1, 2025, and all questions were resolved in July 2025. Participants who made correct predictions earned coins, and the top performers were awarded monetary prizes. 

The final resolutions were made by a panel of four expert judges, each with over seven years of professional experience in software engineering, AI/ML research, and product development at the company. Judges did not participate in the initial preparation of the features for prediction. Before resolution, the organizing team compiled a summary report~\cite{supp_mat} on the market status of each feature based on analysis of the feature set of \company' IDEs, and 20 other tools that were considered, e.g., Windsurf, Cursor, GitHub Copilot. This document served as a reference for the judges, who reached each final decision through several rounds of structured discussion.

\subsubsection{\textbf{Data analysis}} We analyzed the prediction market results along three dimensions: participant engagement, accuracy of forecasts, and final probability estimates. Engagement was measured using two metrics: trading volume and market liquidity. Trading \textit{Volume} captures the total number of virtual coins bet across both outcomes of a question, reflecting overall activity. Market \textit{Liquidity} reflects the breadth of participation and was calculated by assigning a base value of 50 to each question, with 20 additional points added for every unique participant who placed a bet. Accuracy was assessed by comparing the community’s majority prediction with the actual outcome, as defined by the resolution criteria. The final probability estimate for each question was calculated as the proportion of coins placed on the YES outcome relative to the total number of coins bet on that question.

\subsection{FINDINGS}

To assess whether internal practitioners correctly anticipate which AI features inspired by user needs are likely to be implemented, we analyzed prediction market outcomes along three dimensions: forecast accuracy, final probability estimates, and participant engagement. The results are summarized in Table~\ref{tab:tournament_results}.

\paragraph{\textbf{Forecasting performance}} Ultimately, experts concluded that 4 of the 9 features have appeared on the market in any coding tool, and 2 of the 9 appeared specifically in \company IDEs by the end of the prediction window (July 2025). The market produced accurate forecasts for 16 of the 18 questions (88\%), showing that at least for short-term predictions the professional community demonstrates considerable precision. 

Two of the incorrect forecasts were false negatives: participants assigned low probabilities to features that were actually released: \textcolor{jbblue}{Project-wide Context Awareness} in \company IDEs, and \textcolor{jbblue}{Business-context Awareness} in any of the tools available on the market. This pattern suggests a potential \textit{\textbf{a blind spot}}: while these needs were voiced by users, they were underestimated by internal practitioners, potentially due to organizational constraints or differing views on feasibility. What is interesting, no feature was overestimated, meaning there were no cases where the market predicted that a feature would be implemented, but it actually was not. This conservative bias suggests that \textit{\textbf{market signals can be trusted to filter out overly speculative ideas}} (e.g., \textcolor{jborange}{Brain-computer Interface Coding}), but may understate the pace of external innovation.

\paragraph{\textbf{Alignment with user needs}} Among the 9 AI features derived from user needs in Phase 1, three were correctly forecast as likely to be implemented and were in fact released: \textcolor{jbgreen}{Multi-file Generation}, \textcolor{jbblue}{Project-wide Context Awareness}, and \textcolor{jbgreen}{Sketch2Code}. These features respond to the needs around \colorbox{jbgreen!50!}{AI in Implementation} and \colorbox{jbblue!50!}{Context-aware Technology}. Practitioners identified them as feasible and valuable, and their accurate prediction indicates \textit{\textbf{strong consensus and implementation readiness}}.

In contrast, features that required deeper system integration or organizational data, such as \textcolor{jbblue}{Proactive Documentation}, \textcolor{jbblue}{Debugging}, \textcolor{jbblue}{Optimization}, and \textcolor{jbgreen}{AI Real-Time Refactoring}, received low probability scores (all below 33\%) and were not released. This suggests that despite user interest, \textit{\textbf{ the technical or infrastructural barriers prevent realization of such features}}. Here, the low probabilities assigned to these features do not imply irrelevance. On the contrary, their presence in the prioritized by experts list of interesting features from Phase 1 and the fact that practitioners did vote for them, albeit cautiously, indicates that these needs remain relevant but unaddressed by current tools. This creates a clear opportunity for future development. 

Our results suggest that prediction markets can help professional communities identify feasible features and areas for innovation, as well as guide product planning and standardization efforts around AI-assisted development.

Future work should investigate intervention strategies to close the recognition-implementation gap. Potential approaches include: (1) \textit{capability reassessment workshops} where development teams re-evaluate the feasibility of previously dismissed features using structured technical analysis; (2) \textit{incremental implementation pathways} that break complex features into manageable development milestones; and (3) \textit{user-developer collaboration sessions} that allow practitioners to better understand the business value and user impact of challenging-to-implement features. Additionally, longitudinal studies could track whether features initially deemed infeasible by practitioners are successfully implemented by competitor tools or through alternative technical approaches, providing evidence for or against the accuracy of initial feasibility judgments.

\begin{RQanswer}
\textbf{Answer to RQ3.} Practitioners correctly predicted the outcome for 16 of the 18 features, showing that collective judgment aligns with both user needs and actual realization. Features related to implementation and context awareness were accurately anticipated and released. 
\end{RQanswer}
\section{DISCUSSION}

This study bridges qualitative insights from developers with internal feasibility assessments, offering a dual-perspective view on in-IDE AI. By integrating user voices and practitioner judgment, we extend prior literature on human–AI experience in software engineering.

While past studies have mapped general expectations for AI tools for coding, our work focuses on in-IDE use, capturing and systematizing nuanced needs shaped by the real-time and context-rich nature of coding environments. Our findings support existing calls for context-aware~\cite{sergeyuk2024using}, non-intrusive~\cite{Mozannar_Bansal_Fourney_Horvitz_2024,Wang2023How}, transparent~\cite{zhou2024exploring}, ethical~\cite{Liu2024Empirical,Mcnutt2023Design}, and customizable~\cite{liang2024large} AI systems. However, our study also reveals new priorities: integration into broader business workflows, support for skill building, and proactive behavior.

The inclusion of Churners and Non-Users helps surface barriers often missed in adoption-centered research~\cite{sergeyuk2025human}. In particular, there is a need for trust-building mechanisms and accessible onboarding.

Importantly, the prediction market reveals a pattern of conservatism: features that are straightforward to implement are correctly anticipated and realized. In contrast, proactive features and those tied to code maintenance are underestimated despite user interest. This points to a gap between user demand and what teams believe is feasible in the near term.

Together, our two-phase approach illustrates not only what developers want and expect from AI in IDEs, but also which of these expectations are likely to shape product development. Based on our research findings, we propose several key recommendations to guide the future development of in-IDE AI assistants to improve usability, enhance trust, and better integrate AI into developers' workflows. These recommendations could be generalized to most IDEs and editors since they represent the requests and needs of the users rather than specific features of a specific system.

\noindent{\colorbox{jbblue!50!}{\textit{\textbf{Technology Improvements.}}}} 
\textbf{Create Proactive AI.} 
All three participant groups expressed a need for Proactive AI, which was also rated as relevant in the prediction market. AI assistants should take a more proactive role in development, moving beyond reactive responses to explicit prompts and recommendations. Key actions include: \textbf{(1)}~\textit{Autonomous suggestions}: continuously analyze the code and suggest improvements, such as identifying potential bugs, performance issues, or optimization opportunities; \textbf{(2)}~\textit{Automated actions}: instead of suggesting changes, the assistant should be able to perform them proactively, e.g., automatically creating new files with boilerplate code or project structure when a user starts a new project.

\noindent\textbf{Build User Trust.} 
Trust is critical for AI adoption in development environments. 
Developers should prioritize features that build transparency and reliability: \textbf{(1)}~\textit{On-premise AI}: offer on-premise AI to enhance security and privacy for organizations with sensitive codebases; \textbf{(2)}~\textit{Transparent outputs}: providing explanations and sources (\textit{e.g.}, Stack Overflow links) for the assistant's suggestions helps users understand and trust the output; \textbf{(3)}~\textit{Up-to-date data}: ensure that the AI assistant leverages current and relevant libraries, frameworks, and coding standards to provide reliable recommendations.

\noindent\textbf{Ensure Accessibility.}
AI technologies should be usable by developers from various regions and backgrounds: \textbf{(1)}~\textit{Flexible pricing and access}: enable users to choose the features they want to pay for and provide options to access AI from various platforms; \textbf{(2)}~\textit{Language support}: incorporate multi-language support and localized interfaces to ensure that AI tools are usable in different countries; \textbf{(3)}~\textit{Inclusive access}: consider regional differences in infrastructure by offering lightweight or offline versions for areas with limited internet connectivity.

\noindent\textbf{Ensure Privacy.}  
Privacy and security were unresolved concerns for many participants. To support adoption under organizational and national constraints, AI systems should: \textbf{(1)}~\textit{Support on-premise and local AI}: allow use without sending code to external servers, enabling companies to keep data within national or internal infrastructure; \textbf{(2)}~\textit{Enable auditability}: ensure that AI-generated code is clearly marked to support inspection and maintain trust.

\noindent\colorbox{jbgreen!50!}{\textit{\textbf{Simplifying Programming Tasks.}}} 
\textbf{Look Beyond Code Writing.} 
To support the needs of all users' groups, AI should not only assist in code writing but also in code-base navigation, evaluation, and in improving code quality: \textbf{(1)}~\textit{Codebase navigation}: support developers in understanding and exploring the codebase through contextual summaries and links between related components; \textbf{(2)}~\textit{Project analysis}: AI assistants should be able to provide evaluations of the codebase and suggest optimizations for performance, security, or maintainability.

\noindent\textbf{Integrate AI Systems.} 
Participants in all three groups expressed an unmet need for in-IDE AI systems that are tightly connected with internal systems and other engineering tools. Smooth integration with external systems will boost the AI's utility throughout the SDLC: \textbf{(1)}~\textit{CI/CD integration}: enable the AI to support continuous integration and continuous deployment pipelines, automating build and deployment tasks; \textbf{(2)}~\textit{Task tracking integration}: connect with task management tools, allowing the AI to monitor issue statuses, update progress, and link IDE activities with external sources; \textbf{(3)}~\textit{Commit management}: automate and manage commits, generating commit messages and simplifying the version control process within IDE; \textbf{(4)}~\textit{Code review support}: integrate with code review platforms to assist in reviewing code, and suggesting improvements without leaving the IDE.

\noindent\colorbox{jborange!50!}{\textbf{\textit{Technology Interaction.}}}
\textbf{Streamline In-IDE Interaction.}  
Participants emphasized the importance of minimizing context switching and ensuring smooth interaction between AI and existing IDE features. To support this: \textbf{(1)}~\textit{Enable embedded interfaces}: keep AI interaction within the IDE through integrated chat and in-line code explanations to reduce disruption; \textbf{(2)}~\textit{Support multimodal input}: allow voice commands or device-based control (e.g., wearables) to increase accessibility and flexibility.

\noindent\colorbox{jbpurple!50!}{\textbf{\textit{Technology Customization.}}}
\textbf{Personalize AI.} 
Developers, especially Churners, seek personalized AI experiences that adapt to their unique needs and workflows. This can be achieved by: \textbf{(1)}~\textit{Custom setups}: allow users to configure the AI assistant, \textit{e.g.}, selecting models, defining interaction methods, or choosing the technology stack; \textbf{(2)}~\textit{Learning from interactions}: let the assistant learn and adjust from user feedback (\textit{e.g.}, declining specific completions).

\noindent\colorbox{jbpink!50!}{\textbf{\textit{Simplifying Skill Building.}}} 
\textbf{Reduce the Learning Curve.}
Onboarding and education are key for users to fully utilize AI assistants. All three groups of participants expressed challenges and needs in this regard: \textbf{(1)}~\textit{User training}: offer tutorials and examples to help users interact effectively with the AI; \textbf{(2)}~\textit{Prompting assistance}: implement prompt suggestions and provide users with feedback on their prompts to improve their AI interaction skills.
\section{THREATS TO VALIDITY}

\textbf{Internal validity.}
The sample size was determined to reach data saturation; therefore, conclusions are drawn based on the patterns observed within this sample, and further research with a larger or different sample may reveal additional insights or variations of existing topics.

The interview guide was developed and piloted to refine the questions and structure. However, despite these efforts, there may still be inherent biases in the questions or the way they were presented that could influence participants' responses. The iterative process of developing and refining the interview guide aimed to mitigate this risk.

The prediction market participants were JetBrains employees whose predictions may reflect insider knowledge. While this limits external generalizability, it strengthens ecological validity as these practitioners directly shape product decisions.

\textbf{External validity.}
The sample was recruited from a pool of developers using \company products. By focusing on \company' users but allowing for different AI tools, we aimed to balance consistency with a broad range of experiences, ensuring that our findings are applicable beyond any single tool. However, it may not fully capture the diversity of experiences and perspectives of developers using different IDEs, potentially limiting the findings' generalizability. Furthermore, our interview sample consists predominantly of senior practitioners, potentially underrepresenting juniors' needs. Future research should recruit across experience levels and IDE ecosystems to better capture variation in AI tool expectations.

Additionally, the interviews and prediction market were conducted during May-June 2024 and December 2024 - June 2025, respectively, a period characterized by rapid evolution in AI-assisted development tools and changing developer attitudes toward AI integration. The findings reflect developer experiences and expectations during this specific temporal context. Given the fast-paced nature of AI tool development and adoption, developers needs, trust levels, and feature priorities may shift as the technology matures and becomes more ubiquitous. Future research should consider how temporal factors influence the stability of the identified personas and feature preferences over time.

\textbf{Construct validity.}
Our thematic analysis followed established qualitative research protocols with iterative coding and consensus-building between two researchers to ensure consistency and reliability. Our goal was to surface previously unexplored differences across adoption stages to inform product strategy, not to establish statistical significance. While our approach cannot statistically validate whether observed thematic patterns represent true group distinctions versus differential reporting rates, this limitation is inherent to exploratory qualitative research with modest sample sizes. The practical value of these insights is demonstrated by their successful translation into actionable features in our prediction market phase, where internal experts and practitioners recognized the feasibility of the features addressing the user needs we identified. Future work could complement our exploratory findings with larger-scale quantitative validation, but the qualitative patterns we uncovered provide essential foundational insights for understanding developer personas in AI tool adoption.

In the second phase, conceptual features derived from interviews were used to formulate prediction market questions. However, these features did not always align well with concrete implementation outcomes. While a structured resolution guide and expert panel were used to mitigate this issue, some interpretive variability may remain, affecting the alignment between predicted outcomes and actual market developments.
\section{CONCLUSION}

This study examined what different groups of developers want and expect from AI in IDEs, focusing on unmet needs and anticipated features. Through interviews with 35 Adopters, Churners, and Non-Users, we surfaced a structured landscape of user expectations across five areas:\textit{Technology Improvements}, \textit{Programming Tasks}, 
\textit{Interaction}, \textit{Customization}, and \textit{Skill Building}.
These highlight the demand for more advanced, reliable, and trustworthy AI experiences within development environments. To assess implementation feasibility, we conducted an internal prediction market with 102 practitioners betting on 9 AI features covering unmet needs. The results show alignment between user needs and practitioner judgment for core implementation and context-aware features. However, proactive behaviors and long-term code maintenance remain both underdelivered and underestimated, pointing to a systemic blind spot. Our findings bridge user expectations and product planning by showing where practitioner foresight aligns, or fails to align, with real-world developer needs. Tool builders should prioritize context-awareness and implementation support today, while investing in proactive and maintenance features for the longer term. Future work should explore how to operationalize unmet users' needs that are recognized but deprioritized by tool builders.

\begin{acks}
This work was supported by JetBrains as part of the AI for Software Engineering (AI4SE) collaboration with Delft University of Technology.
\end{acks}

\bibliographystyle{ACM-Reference-Format}
\bibliography{main_refs}

\end{document}